\documentclass[%
 reprint,
superscriptaddress,
 amsmath,amssymb,
 aps,
]{revtex4-2}

\usepackage{graphicx}
\usepackage{dcolumn}
\usepackage{bm}
\usepackage[version=3]{mhchem} 
\usepackage{xcolor} 
\usepackage[utf8]{inputenc}

 \usepackage{ulem}

\AtBeginDocument{\normalem} 

\begin{document}

\preprint{APS/123-QED}
\title{
Attosecond-resolved probing of recolliding electron wave packets \\ in liquids and aqueous solutions}

\author{Angana Mondal}
\affiliation{Laboratorium für Physikalische Chemie, ETH Zürich, Zurich, Switzerland}
\author{Nicolas Tancogne-Dejean}
\email[]{nicolas.tancogne-dejean@mpsd.mpg.de}
\affiliation{Max Planck Institute for the Structure and Dynamics of Matter, Luruper Chaussee 149, 22761 Hamburg, Germany}
\affiliation{Research Center Future Energy Materials and Systems of the University Alliance Ruhr and Interdisciplinary Centre for Advanced Materials Simulation, Faculty of Physics and Astronomy, Ruhr University Bochum, Universitätsstraße 150, D-44801 Bochum, Germany }
\author{George Trenins}
\affiliation{Center for Free-Electron Laser Science CFEL, Deutsches Elektronen-Synchrotron DESY, Notkestra\ss e 85, 22607 Hamburg, Germany}
\author{Sona Achotian}
\affiliation{Laboratorium für Physikalische Chemie, ETH Zürich, Zurich, Switzerland}
\author{Meng Han}
\affiliation{Laboratorium für Physikalische Chemie, ETH Zürich, Zurich, Switzerland}
\affiliation{J. R. Macdonald Laboratory, Kansas State University, Manhattan, Kansas, 66506, USA}
\author{Tadas Balciunas}
\affiliation{Laboratorium für Physikalische Chemie, ETH Zürich, Zurich, Switzerland}
\author{Zhong Yin}
\affiliation{Laboratorium für Physikalische Chemie, ETH Zürich, Zurich, Switzerland}
\affiliation{International Center for Synchrotron Radiation Innovation Smart, Tohoku University, Sendai, Japan}
\author{Angel Rubio}
\affiliation{Center for Free-Electron Laser Science CFEL, Deutsches Elektronen-Synchrotron DESY, Notkestra\ss e 85, 22607 Hamburg, Germany}
\affiliation{Physics Department, University of Hamburg, Luruper Chaussee 149, 22761 Hamburg, Germany}
\affiliation{Max Planck Institute for the Structure and Dynamics of Matter, Luruper Chaussee 149, 22761 Hamburg, Germany}
\affiliation{The Hamburg Centre for Ultrafast Imaging, Luruper Chaussee 149, 22761 Hamburg, Germany}
\affiliation{Center for Computational Quantum Physics (CCQ), The Flatiron Institute, 162 Fifth Avenue, New York NY 10010, USA}
\author{Mariana Rossi}
\affiliation{Center for Free-Electron Laser Science CFEL, Deutsches Elektronen-Synchrotron DESY, Notkestra\ss e 85, 22607 Hamburg, Germany}
\author{Ofer Neufeld}
\email[]{ofern@technion.ac.il}
\affiliation{Technion Israel Institute of Technology, Schulich Faculty of Chemistry, Haifa 3200003, Israel}
\author{Hans Jakob W{\"o}rner}
\email[]{hwoerner@ethz.ch}
\affiliation{Laboratorium für Physikalische Chemie, ETH Zürich, Zurich, Switzerland}

\begin{abstract}
High-harmonic spectroscopy (HHS) in liquids promises real-time access to ultrafast electronic dynamics in the native environment of chemical and biological processes. While electron recollision has been established as the dominant mechanism of high-harmonic generation (HHG) in liquids, resolving the underlying electron dynamics has remained elusive. Here we demonstrate attosecond-resolved measurements of recolliding electron wave packets, extending HHS from neat liquids to aqueous solutions. Using phase-controlled two-colour fields, we observe a linear scaling of the two-colour delay that maximizes even-harmonic emission with photon energy, yielding slopes of 208$\pm$55 as/eV in ethanol and 124$\pm$42 as/eV in water, the latter matching ab initio simulations (125$\pm$48 as/eV). In aqueous salt solutions, we uncover interference minima whose appearance depends on solute type and concentration, arising from destructive interference between solute and solvent emission. By measuring the relative phase of solvent and solute HHG, we retrieve a variation of electron transit time by 113$\pm$32 as/eV, consistent with our neat-liquid results. These findings establish HHS as a powerful attosecond-resolved probe of electron dynamics in disordered media, opening transformative opportunities for studying ultrafast processes such as energy transfer, charge migration, and proton dynamics in liquids and solutions.
\end{abstract}

\maketitle


\section*{Introduction}

\noindent 
Electron dynamics in liquids underpin fundamental processes ranging from radiation-induced damage to photocatalysis and prebiotic molecule formation \cite{Elahe2015,Messina2013,Maiuri2020,Lan2024,Naumann2017,He2025,Sajeev2023, Wu2024}. While these phenomena unfold on femtosecond timescales, accessing the earliest, attosecond-scale electron motion is essential to understanding how chemical reactivity, charge transfer, and energy redistribution are initiated in complex environments \cite{jordan20a,gong22a,gong22b,woerner24a,li24a,Woerner2025}. Over the past two decades, attosecond spectroscopy has revolutionized our understanding of electron dynamics in gases and solids, providing insights into tunneling delays, orbital reshaping, and band-structure evolution in strong laser fields \cite{ghimire11a,shafir12a,kraus15b,vampa15a,luu15a}.
These breakthroughs have been enabled primarily by HHS -- a technique that encodes attosecond-resolved temporal information into the phase and amplitude of the emitted radiation. In gases, this arises from the recollision of a liberated electron with its parent ion, providing insights into tunneling, ionization, and recombination dynamics \cite{dudovich06a, shafir12a, Pedatzur2015}. In solids, by contrast, HHG stems from interband recombination and intraband current modulation, allowing access to band dynamics, phase transitions, and carrier scattering processes in strong fields \cite{ghimire11a,vampa15a,luu15a,Hohenleutner2015,bauer18a,luu18b,you17a,
ghimire19a,Bionta2021,Alcala2022,uchida22a,yue22a,neufeld23a,chang24a,kim25a}. Unambiguous determination of the harmonic phase and its mapping to the different components of the electron motion, namely ionization, propagation and recombination, is central towards reconstructing ultrafast electron dynamics in quantum systems.\\
\begin{figure*} [t!]
\includegraphics[width=0.9\textwidth]{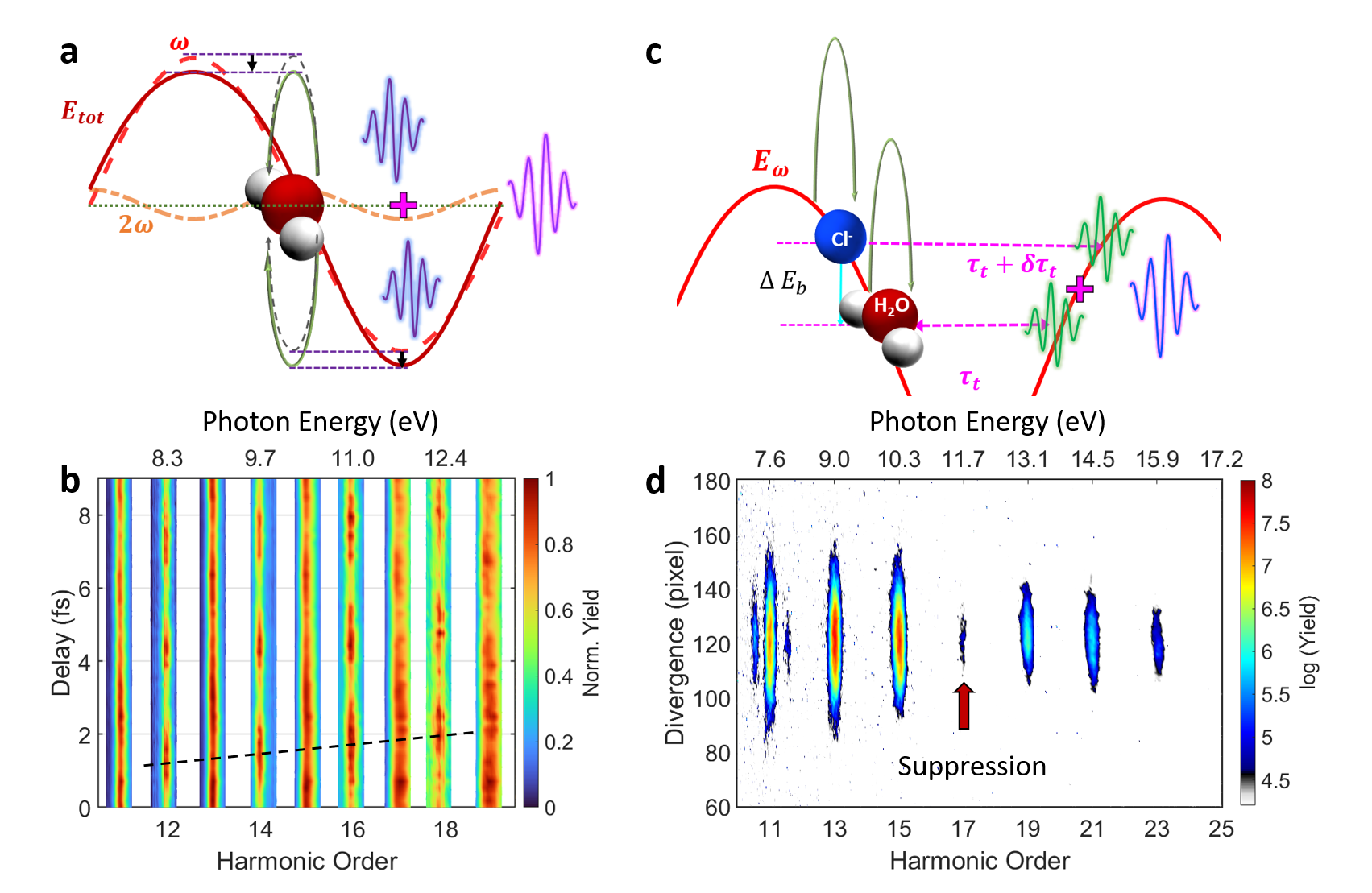}
\caption{
\textbf{Active and passive attosecond interferometry of liquids.} 
\textbf{a} Active interferometry consists in driving the HHG process with a phase-controlled two-colour ($\omega$/2$\omega$) laser field. This causes a different excursion of the laser-driven electron wave packet in consecutive half cycles, which translates into the emission of even harmonics that depend on the phase delay between the two colours.
\textbf{b} Intensity of high harmonics generated in liquid ethanol with 1800/900-nm two-colour laser pulses of 59 fs/42 fs duration, focused to a peak intensity of 2.4 $\times$10$^{13}$ W/cm$^2$/1.8 $\times$10$^{10}$ W/cm$^2$. The black line connects the maxima of even harmonic orders.
\textbf{c} Passive interferometry consists in driving HHG in a salt solution. This leads to high-harmonic emission from the solvent molecules and the solvated anions, whereas emission from the solvated cations remains negligible. Their different binding energies ($E_b$) result in different excursions, which translate into a phase shift of the emitted harmonics that depends on the difference of binding energies $\Delta E_{\rm b}$.
\textbf{d} High-harmonic spectrum of a 0.8-M solution of NaCl in water driven by a 1800-nm, 3.3 $\times$10$^{13}$ W/cm$^2$ laser pulse. The emission of H17 is strongly suppressed as a consequence of destructive interference between the emissions of water and Cl$^-$.
}
\label{fig:one}
\end{figure*}
Yet, despite its promise, obtaining time-resolved information from liquid-phase HHS has remained elusive, with the link between the emitted photon energy and the characteristic time scales of the electron dynamics in the liquid phase unexplored to date. This limitation arises from the unique challenges posed by liquids: high density, strong optical absorption, electron scattering and lack of long-range order, all of which make ultrafast measurements and computations demanding. While ultrafast techniques such as photoelectron spectroscopy and transient X-ray absorption spectroscopy have been extended to bulk liquids to investigate electronic and molecular dynamics \cite{Suzuki2019,Woerner2025}, attosecond time-resolved measurements have so far remained restricted to a measurement of the photoionization delays of liquid- vs. gas-phase water \cite{jordan20a,woerner24a} and a single pump-probe delay of X-ray transient absorption \cite{li24a}.\\

Recent experiments have successfully extended HHG to the liquid phase \cite{dichiara09a,luu18a, yin20a, Svoboda21, mondal23a, Alexander2023,yang24a}, identifying electron recollision as a viable mechanism in bulk liquids \cite{mondal23a,mondal2023few,zeng20a,xia22a,chen23a}. Unlike in gases, where electron quantum paths are dominated by laser parameters, the maximum quantum path length in liquids at ambient pressures is limited primarily by the intrinsic mean free path (MFP) of low-energy electrons and is largely independent of laser wavelength, intensity, or pulse duration \cite{mondal23a,mondal2023few}. 
However, attoseond-resolved HHS in liquid phase has not been achieved to date, and it has remained unclear how the timing of the recolliding electron wave packets could be measured.\\

In this work, we introduce a new paradigm for attosecond-resolved spectroscopy of liquids and aqueous solutions. We demonstrate all-optical attosecond interferometry in the liquid phase using two complementary, yet independent, approaches:
\begin{enumerate}
    \item \textbf{Active interferometry}, based on a phase-controlled two-colour ($\omega$/2$\omega$) field \cite{dudovich06a}, which determines the delay that maximizes symmetry breaking between consecutive half cycles for each emitted photon energy (as illustrated in Fig. \ref{fig:one}a);
    \item \textbf{Passive interferometry}, exploiting spectral interference between HHG from solute and solvent species in salt solutions to extract the electron transit times (as illustrated in Fig. \ref{fig:one}c).
\end{enumerate}

These methods offer two complementary views on the attosecond dynamics of electron dynamics in liquids with $\sim$30 as/eV precision. 
Remarkably, these two techniques provide independent and matching verifications of the characteristic timescales of ultrafast electron dynamics in solutions, each with its own advantages.
Rather than being coincidental, we argue that this agreement reflects a weak dependence of the ionization times on the emitted photon energy, in agreement with the quantum-orbit description of HHG.
By establishing a mapping of emitted photon energy to the attosecond dynamics of the electronic wavepacket in chemically relevant environments under ambient conditions, this work expands the domain of attosecond HHS into uncharted, but highly relevant regimes. 
The ability to track charge motion with sub-cycle precision in liquids lays the foundation for future studies of charge migration in solution, ultrafast solvation dynamics, and light-induced control of liquid-phase chemical reactions, with potentially far-reaching implications for photochemistry, biophysics, and energy science.
\section*{Results}
We first discuss the two interferometric techniques, starting with Fig. \ref{fig:one}a that illustrates active interferometry. This method relies on HHG driven by a phase-controlled two-colour laser field, composed of a fundamental frequency $\omega$ and a weak second-harmonic (2$\omega$) component. In contrast to a single-colour laser field, where electron quantum paths in consecutive half cycles of the electric field are symmetric and only odd harmonics are emitted \cite{ben93a}, the addition of a weak 2$\omega$ component breaks this symmetry and leads to the generation of even harmonics with an intensity that oscillates periodically with the phase delay between the two colours.
Owing to the different phases accumulated by the electron wave packet in two consecutive half cycles, each even harmonic maximizes when the amount of symmetry breaking along its quantum path is maximal -- encoding information about the emission time of the harmonics into the two-colour delay at which the even harmonic maximizes \cite{dudovich06a,vampa15a}.

As shown in Fig. \ref{fig:one}b, we observe clear modulations of all even harmonics, whereby the maxima shift to larger delays as a function of the emitted photon energy $\Omega$. This indicates that higher harmonic orders are emitted later than lower ones, rather than being synchronized. In other words, it reveals that high-harmonic emission from liquids is positively chirped.

In contrast to active interferometry, passive interferometry measures the phase difference between harmonic emissions from two species with different $E_{\rm b}$'s. Figure \ref{fig:one}c illustrates this concept. In a solution, the different $E_{\rm b}$'s of solute and solvent lead to different electron transit times in the laser field. As a result, the emitted harmonics from each species acquire distinct phases. When the resulting phase difference, $\Delta \phi$, is close to an odd multiple of $\pi$, destructive interference occurs at that harmonic order. In atoms and molecules, the phase shift between harmonics emitted by species with different $E_{\rm b}$'s follows the relation $\Delta \phi = \Delta E_b*\tau (\Omega)+\delta\phi$, where $\Delta E_{\rm b}$ is the binding-energy difference, $\tau (\Omega)$ is the transit time of electron wavepackets contributing to the photon energy $\Omega$ \cite{kanai07a} and $\delta\phi$ is the dipole phase difference of the two emitting species \cite{woerner10a,woerner10b,woerner10c,azoury19a}. Here, we demonstrate that this principle can be extended to bulk liquids and aqueous solutions within the first plateau (whereas a different result is expected in recently observed second-plateau harmonics that arise from a distinct physical mechanism \cite{mondal25a}). Figure \ref{fig:one}d shows that the destructive interference condition is fulfilled for a 0.8-M solution of NaCl in water, driven by an 1800 nm field of a peak intensity 2.6$\times$10$^{13}$ W/cm$^2$. This enables us to probe how the harmonic phase varies with photon energy and extract attosecond time-resolved information about the underlying electron dynamics.

With the conceptual framework in place, we next examine what each interferometric approach reveals about attosecond electron dynamics in liquids.

\subsection*{Active interferometry: Two-colour field modulation and recollision timing in liquids}

\begin{figure*} [t!]
\includegraphics[width=\textwidth]{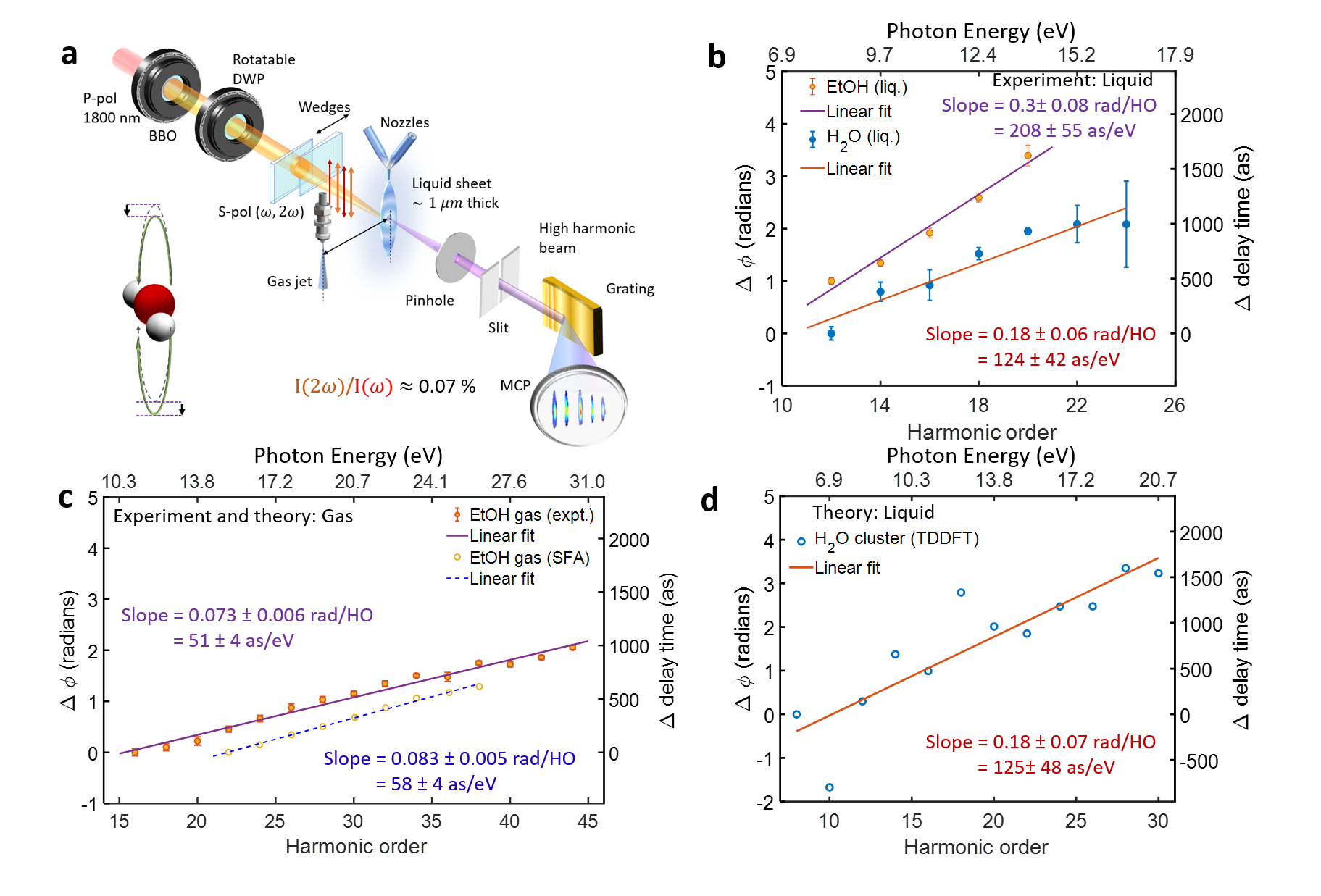}
\caption{
\textbf{Active attosecond interferometry of liquids.} 
{\textbf{a} Schematic of the experimental setup. Laser pulses centered at 1800 nm and a weak ($\sim$ 0.07 \% in intensity) parallel-polarized second harmonic are focused on the flat liquid jet, in which they generate both even and odd harmonics. The delay between the fundamental and second harmonic is controlled by a pair of fused-silica wedges. The generated high harmonics pass through a slit onto a grating that disperses the harmonic orders, which are then detected using an microchannel plate (MCP) coupled to a phosphor screen and a camera. A heatable bubbler setup is mounted on the same manipulator as the flat jet at a distance of 2.5 cm from its center for measuring gas-phase HHG after a lateral translation of the manipulator. 
\textbf{b} Relative phase of the intensity oscillations of even harmonics emitted from liquid H$_2$O (blue circles) and liquid ethanol (yellow circles) at peak fundamental intensities of 3.6$\times$10$^{13}$ W/cm$^2$ and 2.4$\times$10$^{13}$ W/cm$^2$, respectively.
\textbf{c} Same as \textbf{b} from an experiment on gas-phase ethanol (red circles) using 1800 nm pulses with a duration of 59~fs and a peak intensity of 3.7$\times$10$^{13}$ W/cm$^2$ and a weak 900 nm component ($\sim$ 0.5\% in intensity of 1800 nm) and calculations based on the strong-field approximation (SFA, yellow circles) using $E_b=10.48$~eV (ethanol). \textbf{d} Same as \textbf{b} from a TDDFT calculation on water clusters using 1800 nm pulses with a duration of 24 fs (FWHM) and a peak intensity of 5$\times$10$^{13}$ W/cm$^2$ with a weak 900 nm component (0.04\% intensity of 1800 nm). Linear regressions and their slopes are shown where applicable.
}}
\label{fig:two}
\end{figure*}

The schematic of the experimental setup is shown in Fig.~\ref{fig:two}a. A phase-controlled two-colour ($\omega$-$2\omega$) laser field with parallel polarizations of the two colours is generated by frequency doubling an 1800-nm fundamental pulse in a $\beta$-barium borate (BBO) crystal. 
The phase delay between the $\omega$ and 2$\omega$ components is controlled using a pair of fused-silica wedges. The high harmonics generated by this two-colour field in a liquid flat jet or an interchangeable gas jet are detected in a flat-field imaging spectrometer.

Figure \ref{fig:one}b shows a typical harmonic spectrum of liquid ethanol as a function of the two-colour delay. Despite the low relative intensity of the 2$\omega$ component (0.07 \%), the symmetry of the HHG process is significantly broken, resulting in the appearance of even harmonics. These harmonics exhibit periodic intensity modulations as a function of the $\omega$-2$\omega$ delay, with a modulation period corresponding to the optical cycle of the 2$\omega$ field. The black-dashed line connects the delay values that maximize the yield of each even harmonic, indicating a linear dependence of the optimal delay on the harmonic order.\\

\noindent We then extract the two-colour delay that maximizes a given harmonic order by Fourier analysis. As shown in Figure \ref{fig:two}b, these delays increase approximately linearly with harmonic order for both liquid water (blue circles) and ethanol (orange circles). A linear fit to these data returns slopes of 208 $\pm$ 55 as/eV for liquid ethanol and 124 $\pm$ 42 as/eV for liquid water. 

\noindent The experimental results are further supported by ab-initio time-dependent density-functional-theory (TDDFT) calculations on large water clusters (see methods section for details), which yield a slope of 125 $\pm$ 48 as/eV (Fig. \ref{fig:two}d). 
\noindent The magnitude of these slopes is much larger than in the gas-phase measurements, performed under comparable conditions. Due to its higher $E_{\rm b}$ and longer electron quantum paths, the gas-phase experiment required higher driving intensity (${\sim5}\times10^{13}$~W/cm$^2$) and a stronger 2$\omega$ component (${\sim 0.5\%}$) to observe a clear modulation of the even harmonics. As shown in Fig. \ref{fig:two}c (yellow circles), the extracted slope ($58 \pm 4$~as/eV) is much smaller than in the liquid-phase measurements. This result agrees with calculations based on the SFA for a model system with $E_{\rm b}=$10.48 eV, matching that of gas-phase ethanol (orange circles in Fig. \ref{fig:two}c). Further experimental and theoretical details are provided in the \textit{Methods} section.

The most striking result of these measurements is the large linear slope, which reflects a large positive attochirp of harmonic emission from liquids. 

To extract additional information about the attosecond electron dynamics, we turn to passive interferometry. By independently extracting the relative phases of high harmonics emitted by different species, we assess whether the large variation of the harmonic emission time observed in active interferometry is a general feature of the liquid-phase HHG mechanism, and whether it persists across different measurement techniques. 
\subsection*{Passive interferometry: Single-colour measurements in salt solutions.}
\begin{figure*} [t!]
\includegraphics[width=\textwidth]{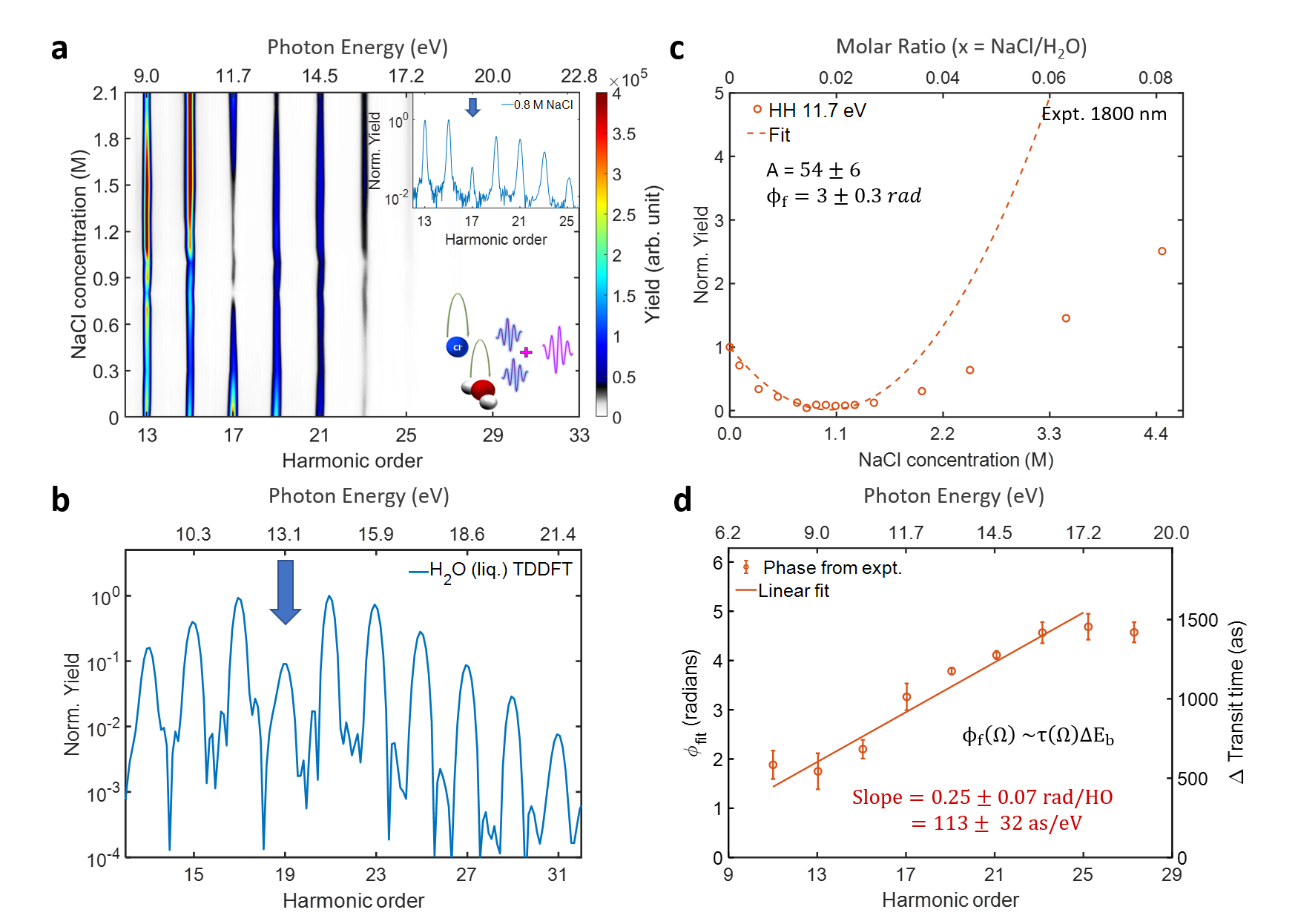}
\caption{
\textbf{Passive attosecond interferometry of salt solutions.} 
{\textbf{a} Dependence of harmonic yields on the concentration of NaCl in water, showing the suppression of H17 (11.7 eV) for concentrations around 1M. The inset shows the harmonic spectrum for a 0.8-M NaCl solution. These data were obtained with 1800-nm pulses focused to a peak intensity of 3.3$\times$10$^{13}$ W/cm$^2$.
\textbf{b} TDDFT calculation of the high-harmonic spectrum of a 1-M NaCl solution in water using an 1800-nm driving pulse with a peak intensity of 2$\times$10$^{13}$ W/cm$^2$.
\textbf{c} Measured yield of H17 (11.7 eV) as a function of NaCl concentration (circles). The orange dashed line is a fit based on a two-species (H$_2$O and Cl$^-$) interference model. 
\textbf{d} Variation of the relative phase of emission from Cl$^-$ and H$_2$O, extracted from a fit to the interference model (left axis) and variation of the electron transit time (right axis), as a function of the emitted photon energy.}}
\label{fig:three}
\end{figure*}

In its simplest implementation, this technique requires two species contributing to the harmonic emission. We realize the two-emitter system with aqueous solutions of NaCl, containing  H$_2$O, hydrated cations (Na$^+_{\rm(aq)}$) and anions (Cl$^-_{\rm(aq)}$), whereby we drop the subscripts ``(aq)" for brevity from hereon. Owing to the significantly higher $E_{\rm b}$ of Na$^+$ (35.4 eV) compared to liquid H$_2$O (11.67 eV) and Cl$^-$ (9.55 eV) \cite{perry20a,Jagoda-Cwiklik2008,Seidel2016}, the cations do not contribute measurably to high-harmonic generation. This statement is confirmed by the experimental observation that the emitted HHG spectra and especially the position of the interference minimum strongly depend on the nature of the anion (Extended Data Fig., EDF 1), but not on that of the cation (EDF 2).
The concentration of Na$^+$ in H$_2$O serves as the control parameter. Note that we use the relative $E_{\rm b}$'s, referenced to the vacuum level, rather than the energy difference between valence bands and the conduction band of the solution, because (i) the $E_{\rm b}$'s are more accurately known than the band gaps and (ii) the relative $E_{\rm b}$'s also reflect the energy difference of the valence bands because of the common conduction band for solute and solvent (see EDF 5d and 6d), in analogy with negatively doped semiconductors \cite{yu19a}.

Figure \ref{fig:three}a shows high-harmonic spectra as a function of the NaCl concentration. A pronounced minimum in the yield of H17 (11.7 eV) appears around 1 M, with the inset for 0.8 M illustrating the maximal destructive interference. The relative yields of all harmonics vary systematically with concentration, consistent with a phase-sensitive interference between the harmonic emissions from H$_2$O and Cl$^-$.

To further demonstrate that this interference reflects a genuine microscopic effect, we performed elaborate ab-initio TDDFT simulations combined with molecular dynamics (MD) of the salt solution (see the Methods section for details), with the results shown in Fig. \ref{fig:three}b. The calculations reproduce the observed minimum (upshifted to the next higher order), confirming that it arises from the microscopic electronic response as opposed to macroscopic phase-matching and propagation effects, which are not included in our TDDFT simulations.
While the ab-initio simulations confirm the microscopic origin of the harmonic suppression, performing a fine-grained concentration scan is computationally expensive and time-consuming, and can become intractable for either very low, or very high, solute concentrations (requiring either very large simulation cells, or high level theory for properly capturing the interaction between the solute ions, respectively). For the interpretation, we therefore introduce an analytical two-emitter interference model (sketched in Fig. \ref{fig:three}a), which offers an intuitive and transparent way of determining the relative phase shifts and emission-amplitude ratios.

To extract the phase shifts between H$_2$O and Cl$^-$ from experimental data, we fit each harmonic yield as a function of salt concentration using the two-emitter model. For a given solution containing $m_1$ moles of H$_2$O and $m_2$ moles of Cl$^-$ ions, the total intensity $I_\mathrm{total}(\Omega)$ at a given photon energy $\Omega$ is treated as the coherent sum of the contributions from each species:
\onecolumngrid
\begin{equation}
\centering
I_{\mathrm{total}}(\Omega)  = m_1^2 I_{\mathrm{H_2O}}(\Omega) + m_2^2 I_{\mathrm{Cl^-}}(\Omega) +2m_1m_2 \sqrt{I_{\mathrm{H_2O}}(\Omega)I_{\mathrm{Cl^-}}(\Omega)} \cos(\phi_{\rm f}(\Omega)),
\label{eq:one}
\end{equation}
\twocolumngrid
\noindent where $I_{\mathrm{H_2O}}(\Omega)$ and $I{_\mathrm{Cl^{-}}}(\Omega)$ are the single-species emission intensities from 1 mol of particles, and $\phi_{\text{f}}$ is their relative phase. 
Dividing this intensity by the known harmonic yield of an identical pure water sample, gives the normalized yield (Eq. \ref{eq:two}),
which displays the expected quadratic dependence on the molar ratio ($x=m_2/m_1$). The parameters $A(\Omega)=\sqrt{(I_\mathrm{Cl^{-}}(\Omega)/I_\mathrm{H_2O}(\Omega))}$ and $\phi_{\rm f}(\Omega)$ are then retrieved from the measured intensities of HHG from solutions of different concentrations.
\onecolumngrid
\begin{eqnarray}
\frac{I_\mathrm{total}(\Omega)}{I_\mathrm{H_2O}(\Omega) m_1^2}  &=& 1 + 
\frac{m_2^2 I_\mathrm{Cl^-}(\Omega)}{m_1^2 I_\mathrm{H_2O}(\Omega)} +\frac{2m_2}{m_1} \sqrt{\frac{I_\mathrm{Cl^-}(\Omega)}{I_\mathrm{H_2O}(\Omega)}} \cos(\phi_{\rm f}(\Omega)) \\ \label{eq:two}
&=& 1+(A(\Omega)x)^2+2xA(\Omega) \cos(\phi_{\rm f}(\Omega)). \nonumber
\end{eqnarray}
\twocolumngrid

\noindent Figure \ref{fig:three}c shows the fit of the two-emitter model (red dashed line) to the normalized yield of harmonic 17 (H17, 11.7 eV). At low concentrations (up to 1.5~M NaCl) the two-emitter model reproduces the data well (Fig.~\ref{fig:three}c), while at higher concentrations, the measured harmonic yields lie below the model prediction.

This deviation likely has several underlying causes. First, as the salt concentration increases, the probability for the electron wave packet to scatter off Na$^+$ or Cl$^-$ ions rises, since both ions have larger electron scattering cross sections than water. This increased scattering reduces the electron’s mean-free path, suppressing the harmonic yield \cite{mondal23a}. Similar effects have been reported for higher-energy electrons ($\sim$ 65 eV), whose mean-free paths decrease with increasing NaI concentrations above 1 M \cite{olivieri16a}. Second, the addition of salt also increasingly modifies the electronic structure of the solution (EDF 3), especially at higher concentrations where it also changes the geometric structure (EDF 4) which can be expected to alter the HHG yield. 

At low concentrations (${\lesssim 1}$~M), the validity of the two-emitter model is further supported by our electronic-structure calculations combined with the MD simulations, shown in EDF 5 and 6. They indicate that the HOMO in salt solutions remains highly localized on individual water molecules and anions (EDF 5). Meanwhile, the LUMO is delocalized over the simulation box. The localization of the HOMO ensures that the harmonic emission can be treated as originating from discrete species with well-defined $E_{\rm b}$'s, consistent with the observed interference behavior and supporting the validity of the method chosen for extracting the harmonic phase shift. The delocalization of the LUMO and higher-lying states validates the treatment of the electron dynamics as taking place in a quasi-continuum. 
The two-emitter model is also supported by experiments in which we systematically varied the anion. Replacing Cl$^-$ with Br$^-$ and I$^-$ featuring $E_{\rm b}$'s of 9.55~eV, 8.8~eV and 7.9~eV, respectively, and thus an increasing $\Delta E_{\rm b}$ relative to water, we observe that the suppressed harmonic order moves to lower photon energies (EDF 1) -- consistent with the fact that lower harmonic orders correspond to the shorter transit times that are necessary to fulfill the destructive-interference condition. Varying the cation has no significant effect, as expected (EDF 2).

Figure \ref{fig:three}d shows the extracted phase $\phi_{\rm f}$ as a function of the harmonic order. Assuming that the dipole phase difference $\delta\phi$ depends weakly on the photon energy, the variation of $\phi_{\rm f}$ can be converted to the variation of the transit time, shown on the right axis of the figure. The assumption of the weak dependence of $\delta\phi$ on the photon energy, which is expected in the absence of (anti-)resonances as confirmed in Ref. \cite{azoury19a}, is validated in EDF 7.
\begin{figure*} [t!]
\includegraphics[width=\textwidth]{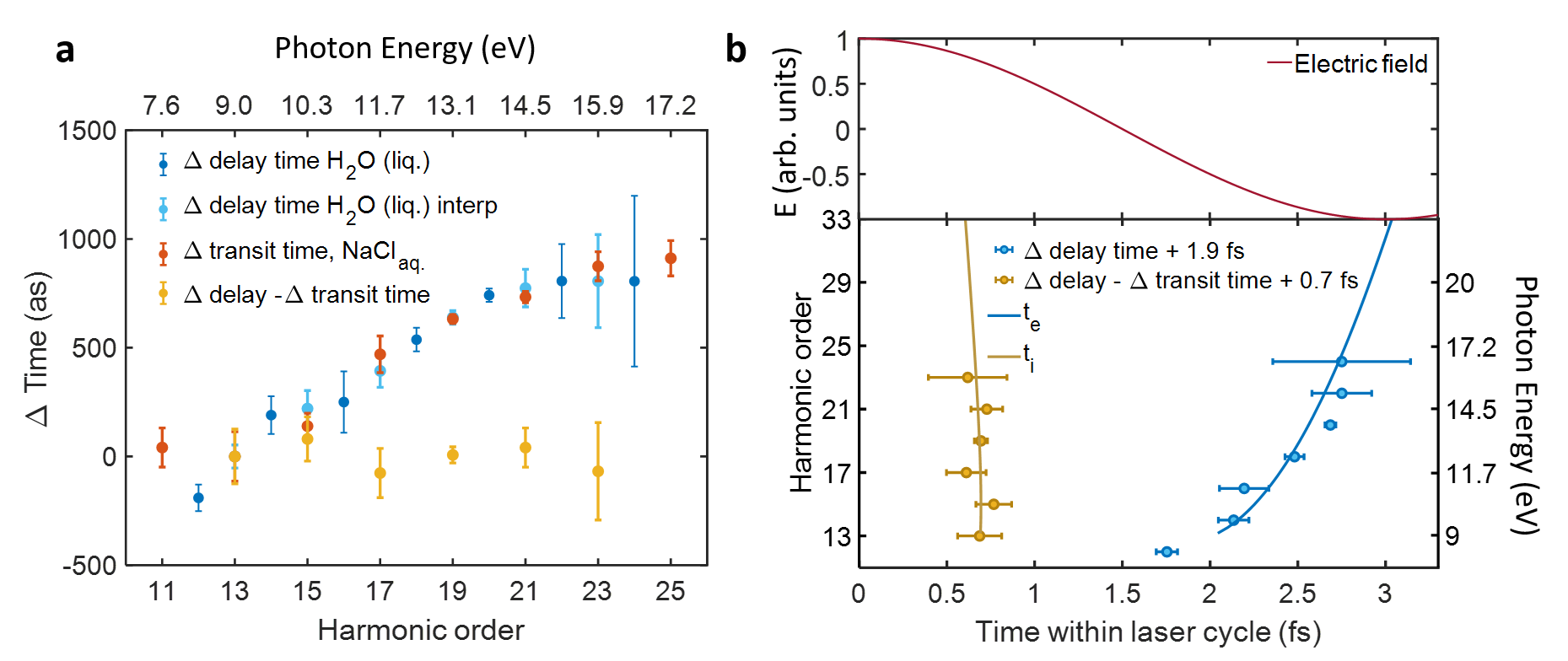}
\caption{\textbf{Attosecond timing of electron dynamics in liquids. a} Variation of electron transit time with photon energy from single-colour measurements on aqueous NaCl solutions (Fig. 3d, red), two-colour delays maximizing even harmonics from measurements on liquid H$_2$O (Fig. 2b, dark blue) and their interpolation onto the odd-harmonic axis (light blue). The difference of the two times, corresponding to the variation of ionization times with photon energy, is shown in yellow.
\textbf{b} Comparison of the extracted variations of ionization (yellow circles) and emission times (blue circles) with a quantum-orbit calculation (lines) described in the Methods Section, at a peak intensity of  3.3$\times$10$^{13}$~W/cm$^2$, for $E_b=$8.83 eV (bandgap of liquid water). The experimental data has been shifted by constants offsets of 0.7 fs (for ionization time, yellow circles) and 1.9 fs (for the emission time, blue circles), to overlay on the curves of the quantum-orbit calculations. These constant shifts were determined from the mean of the difference of the experimental and quantum-orbit results. The driving wavelength used in all experiments and calculations is 1800~nm.}
\label{fig:five}
\end{figure*}

A comparison of the obtained variation of transit times (Fig.~\ref{fig:three}d) with the results of the two-colour measurements on liquid water (Fig.~\ref{fig:two}b), shown in Fig.~\ref{fig:five}a, demonstrates that the determined time intervals agree within the respective experimental errors. This agreement is remarkable, especially for a dynamical system as complex as liquid water, and reflects the attosecond dynamics of electron recollision. 

\section*{Discussion and conclusions}
The attosecond electron dynamics underlying HHG in liquids can be described by three characteristic times: the ionization time ($t_i$), the transit time ($\tau$), and the harmonic emission time ($t_e$). Each of them is encoded in the phases measured in the present work. As explained above, the relative phase $\phi_{\rm f}$ of the harmonic emission from H$_2$O and Cl$^-$ gives access to the variation of the transit time $\tau=t_e-t_i$ and the two-colour delays give access to the variation of $t_e$ \cite{dudovich06a} with photon energy. 
Comparing the variations of $\tau$ and $t_e$ in Fig.~\ref{fig:five}a, we find that they agree within the experimental error. This, in turn, indicates a negligible dependence of the ionization times $t_i$ on the harmonic order. This conclusion is remarkable in that it agrees with the results from gas-phase experiments and the quantum-orbit analysis of gas-phase HHG \cite{shafir12a}, which both demonstrate that the variation of the ionization time with harmonic order is negligible compared to the variation of the emission time with harmonic order.

This experimental observation suggests to attempt a comparison of the measured variations of $t_i$ and $t_e$ with a quantum-orbit calculation that uses the band gap of liquid water as the $E_{\rm b}$.
Figure~\ref{fig:five}b shows the results of such a calculation (full lines), superimposed with the experimental data -- shifted horizontally since only their variation with photon energy is accessible in the experiment. Strikingly, the variations of both $t_i$ and $t_e$ agree with the quantum-orbit calculations within the experimental errors. 
Our results have a number of implications. \\
First, the agreement between active and passive interferometry validates both techniques. It reveals that high-harmonic emission from liquids displays a positive linear chirp, which encodes the attosecond dynamics of the underlying electron quantum dynamics. The common energy dependence of both emission and transit times implies a negligible variation of the ionization times with harmonic order, consistent with the quantum-orbit description of HHG.\\
Second, they establish a quantitative link between the emitted photon energies and the attosecond-scale variation of the recombination times of the electron wave packets with photon energy in the liquid phase. This link turns liquid-phase HHS into a technique with attosecond temporal resolution, capable of resolving electron dynamics on their natural timescale. \\
Third, our measurements demonstrate that high-harmonic emission from solvent and solute molecules is fully coherent at low concentrations (up to $\sim$1-M NaCl or NaBr) within the sensitivity of our measurements and can be used to retrieve the variation of electron transit times on attosecond scales.

Taken together, these results demonstrate all-optical attosecond spectroscopy of electronic processes in liquids and solutions, which can be developed further to study electron tunneling, charge- and energy-transfer processes and electronic dynamics underlying radiation damage, nucleation, proton transfer or phase transitions, as well as electronic processes induced by strong electromagnetic fields.

\section*{Methods}
\subsection{Experimental Methods}
The experiments were carried out at the fundamental laser wavelength of 1800 nm, generated using an optical parametric amplifier (HE-TOPAS) pumped by a 800-nm, 1-kHz titanium-sapphire beam with a pulse energy of $\sim$6 mJ. For the two-colour experiments, the fundamental beam centred at 1800 nm ($\omega$) of pulse energy 400~µJ and pulse duration $\sim$ 55~fs, was used to generate a second harmonic centred at 900~nm (2$\omega$) using a 75~µm $\beta$-barium borate (BBO) crystal. The pulse energy ratio between the $\omega$ and the 2$\omega$ was varied from 0.03\% to 0.7\% by tilting the BBO with respect to the polarization axis of the 1800-nm beam. A dual wave plate was used to align the $\omega$ and 2$\omega$ polarizations, and their relative delay was adjusted via a pair of fused-silica wedges. The beams were then focused onto a liquid flat-sheet target, and the emitted high harmonics were detected with a flat-field imaging spectrometer. The flat jet is created by colliding two cylindrical jets created from two cylindrical nozzles of inner diameters $\sim$54~$\mu$m, resulting in a thickness of $\sim$1~$\mu$m \cite{luu18a,yin20a,chang22a}.
A gas target was mounted on the same manipulator as the liquid jet, to facilitate gas-phase measurement by mere translation of the stage.\\
To ensure that the experiment operates in the perturbative regime of the 900-nm component, we verified two conditions: (1) the absence of higher-order components (e.g., 4$\omega$) in the modulation of the even-harmonic yield, and (2) the out-of-phase oscillations of the even- and odd-harmonics intensities.\\
The 2$\omega$-phase component of the even-harmonic modulation was extracted using Fourier-transform analysis. For each harmonic, the intensity modulation as a function of the $\omega$–2$\omega$  delay was Fourier transformed, and the phase of the 2$\omega$ peak was extracted from the Fourier spectrum.\\
For the concentration scan up to 4-5 M salt concentrations, larger nozzle diameters of $\sim$ 100 $\mu$m were used, to prevent clogging of the jet. However, as the HHG is generated only from the last few 100 nm \cite{luu18a,mondal23a},  a thicker flat-jet $\sim$2-3~$\mu$m, does not significantly modify the spectral features of the emitted HHG. Data recorded with both $\sim$100~$\mu$m nozzles (Fig. 3a) and $\sim$54~$\mu$m nozzles (Fig. 4b), show the same selective suppression of the 11.7-eV, H17 at 0.8 M NaCl.

\subsection{Theoretical Methods}
\textbf{TDDFT cluster calculations.} We performed ab-initio TDDFT simulations for the case of liquids driven by intense $\omega-2\omega$ laser pulses. These simulations followed the approach extensively developed and described in Ref. \cite{Ofer2021}. We employed a 54-water-molecule cluster \cite{54watercluster} that was driven by intense laser pulses as detailed in the main text. The induced dipole response was extracted by solving the time-dependent Kohn-Sham (KS) TDDFT equations of motion employing the Perdew-Burke-Ernzerhof (PBE) \cite{PBEXC} exchange-correlation (XC) functional for the ground state, while neglecting electron-electron interactions (employing the independent particle approximation (IPA)) in the time-dependent simulations, and artificially removing the response of surface-localized states (see details in Ref. \cite{Ofer2021}). We assumed clamped nuclei, which should be applicable up to tens of femtosecond timescales \cite{MengJPCL}. Simulations employed a spherically-shaped real-space grid with cartesian coordinates of spacing 0.4 Bohr utilizing the Octopus code \cite{Tancogne-Dejean2020}, and a time-step of 1.9 as for the temporal propagation. Further, we employed a complex absorbing potential to remove reflections from the boundaries similar to the approach in Ref. \cite{Ofer2021}. From the time-dependent KS states we extracted the dipolar response, $\textbf{p}(t)=\int \textbf{r} n(\textbf{r},t)d^3\textbf{r}$ (with $n(\textbf{r},t)$ the time-dependent electron density) and the dipole acceleration, $\textbf{a}(t)=\ddot{\textbf{p}}(t)$. The HHG spectrum was calculated by Fourier transforming the dipole acceleration with an added super-gaussian temporal window function, $f(t)$: $I_{HHG}(\Omega)=|\int \textbf{a}(t)f(t) e^{-i\Omega t}dt|^2$. Simulations were performed for $\omega-2\omega$ driving fields with varying values of the relative two-colour phase in steps of 0.05$\pi$. In each case the orientation of the cluster was further angularly-averaged to ensure a limit of an isotropic liquid medium (with 12 orientations averaged per pulse, employing additional symmetries of the laser pulse to perform Euler trapezoidal angular integration, as detailed in refs. \cite{Ofer2021,Neufeld2024bench}). We directly averaged the dipole acceleration in each case such that the actual expression from which the HHG spectra was calculated was of the form $\textbf{a}(t)=\int \textbf{a}_\Omega(t)d\Omega$, with $\Omega$ labeling the angular orientation and the integral performed over the Euler angles. Note that the optical gap obtained in the cluster case of liquid water at the PBE level (${\sim 6.8}$~eV) is smaller than the experimental value (at ${\sim 8.8}$~eV), but larger than the value obtained in the supercell simulations described below (at ${\sim 4.5}$~eV, since we artificially removed surface states which slightly increase the gap). From the calculated HHG responses, we extracted the two-colour phase dependence of the harmonic integrated yields, as discussed in the main text, and with an approach identical to the experimental scheme (i.e., directly from the numerical data without further assumptions). This was achieved by Fourier transforming the two-colour phase-dependent harmonic integrated yields to obtain the phase-of-the-phase, which was unwrapped over the even harmonics. 

\textbf{MD simulations.} To obtain a selection of atomic positions (snapshots) for the TDDFT supercell calculations described below, we performed thermal sampling of the machine learning potential trained by O'Neill and co-workers~\cite{oneillCrumblingCrystals2024,oneillPairNot2024} on DFT forces and energies computed with the dispersion-corrected revPBE-D3 functional~\cite{zhangCommentGeneralized1998,grimmeConsistentAccurate2010}. The dynamics were propagated in \mbox{i-PI}~\cite{litmanIPI302024}, using a stochastic velocity rescaling thermostat~\cite{Bussi2007} to equilibrate to a temperature of 300~K. In order to capture nuclear quantum effects in the thermal distribution, we continued sampling using the path-integral molecular dynamics (PIMD)~\cite{Chandler1981} of 12-bead ring polymers coupled to a PIGLET thermostat~\cite{Ceriotti2012}. The HHG calculations are highly sensitive to the intermolecular structure, and \mbox{revPBE-D3} slightly overbinds the liquid~\cite{marsalekQuantumDynamics2017}. To compensate for this effect, we considered PIMD simulations at a selection of densities, namely 0.90, 0.95, 1.00, and 1.10~g/cm$^3$. Overall, HHG spectra derived from snapshots of pure water at 0.90~g/cm$^3$ corresponded best to the experimental signal.
In light of this, production PIMD simulations of aqueous NaCl solutions at molar concentrations of 0~M (111 \ce{H2O} molecules in the simulation box), 1~M (97 \ce{H2O}, 2 NaCl), 3~M (92 \ce{H2O}, 6 NaCl), and 4~M (89 \ce{H2O}, 8 NaCl) were performed at the densities of 0.900~g/cm$^3$, 0.933~g/cm$^3$, 
1.000~g/cm$^3$, and 1.032~g/cm$^3$, respectively, corresponding to experimental densities at 300~K~\cite{pitzerThermodynamicProperties1984}, scaled by a factor of 0.9. For every salt concentration, 50 snapshots were recorded at 5~ps intervals and subsequently fed into the HHG calculations.

\textbf{Ground-state DFT calculations.}
We analyzed the electronic density of states (EDOS) and the band structures of the salt solutions, performing ground-state electronic structure calculations on the simulation snapshots using density-functional theory with the PBE~\cite{PBEXC} and the PBE0~\cite{perdewRationaleMixing1996} exchange-correlation functionals. The calculations were performed using the FHI-aims software package~\cite{blumInitioMolecular2009,renResolutionofidentityApproach2012,abbottRoadmapAdvancements2025},
with the default \textit{tight} settings for the basis sets and real-space quadrature grids. The EDOSs were computed at the $\Gamma$-point. The total and species-projected EDOSs, obtained with PBE and PBE0 and averaged over all snapshots, are shown in EDF~3. The HOMO and LUMO orbitals, as well as the band structures for selected individual snapshots are shown in EDF~5 (EDF~6) using the PBE0 (PBE) functionals.

\textbf{TDDFT supercell simulations.} Complementary to the cluster simulations of liquid HHG, we employed a secondary ab-initio approach using a supercell of liquid water with $\sim$100 molecules in a cubic unit cell (depending on the salt concentration the number of molecules slightly differs). The approach is similar in spirit to simulations in refs. \cite{nourbakhsh2022,MengJPCL}, but employing the MD snapshots derived with the method described above, and with some minor differences. We assumed clamped nuclei during the interaction with the laser pulse, which should be valid on fs timescales (though longer timescale picosecond dynamics and structural rearrangement were considered by averaging over multiple liquid snapshots obtained from the MD simulations as is detailed below). The main logic of employing this approach, that is complementary to the cluster simulations, is that the cluster approach cannot faithfully describe HHG from salt solutions, because the cluster geometry is adopted from previous work, and the clusters are too small to describe a bulk solution. Moreover, multiple MD snapshots are required to obtain a proper electronic structure in such systems \cite{Prendergast2005}. In that respect, the supercell scheme is expected to be more accurate than the less costly cluster simulations (however, we note that due to the high numerical cost of this approach requiring averaging over many MD snapshots and employing large supercells it cannot be employed for simulating the $\omega-2\omega$ experiments).

With this in mind, simulations in the main text that involve HHG for different salt concentrations in water were performed by solving the TDDFT KS equations of motion fully (for all states in the system) in a periodic cubic supercell, using the Octopus code \cite{Tancogne-Dejean2020}, and employing a real-space grid with $\Gamma$-point only \textit{k}-space integration and 0.4 Bohr spatial intervals. We found that in the bulk, likely due to the high laser power and the infinite nature of the periodic system, the IPA is inapplicable. These simulations similarly involve XC at the PBE level (similar level of theory to the cluster case), but where the geometries of the solutions were obtained as discussed above from MD simulations (with a different more accurate XC functional). At each supercell geometry (denoted snapshot, arising from the MD dynamical evolution of the liquid over picosecond timescales) the time-dependent current was extracted, $\textbf{j}_{\Omega,n}(t)$ (with $\Omega$ labeling the angular orientation, and $n$ the snapshot geometry). We averaged the current evaluated from a total of six simulations at varying spatial orientations (different $\Omega$) of the laser for a given snapshot ($\Omega = \pm x,\pm y,\pm z$), and further for 50 snapshots (values $m\in[0,50]$) separated by 5~ps each for a given solution concentration. From the averaged value of the current, $\textbf{J}(t)=\sum_{m}\int \textbf{j}_{\Omega,m}(t)d\Omega$, we computed the HHG spectrum by taking one time derivative and Fourier transforming with a temporal super gaussian filter $f(t)$: $I_{HHG}(\Omega)=|\int \partial_t{\textbf{J}(t)}f(t) e^{-i\Omega t}dt|^2$. This procedure converged the HHG yield at the first plateau region up to $\sim$19~eV and provided an isotropic odd-only harmonic response along the laser polarization axis, as expected from optical selection rules \cite{Neufeld2019Floq}. Further note that slightly lower laser intensities were employed in the bulk supercell to account for dielectric function changes that need not be taken into account in the finite cluster system. This reduced laser peak powers in actual simulations by a factor of $\sim$2.5. We employed simulations in supercells at a relative liquid density of 0.9 compared to standard density at atmospheric pressure and room temperature, because we noted that in typical conditions simulations did not reproduce the sharp harmonic suppression measured experimentally in Fig. \ref{fig:three}. Based on this, we suggest that in the experimental geometry, the density of the liquid is likely lower than that at equilibrium due to the laser irradiation itself. 

\textbf{Quantum-orbit calculations with saddle-point approximation} First, we solved the saddle-point equations in SFA with the approach introduced in \cite{le2016strong}. With the saddle-point approximation, one can easily separate the long and short quantum paths, which have the opposite attochirp. In our calculations we only considered the short quantum paths due to their preferred phase matching. The calculated ionization and recollision times are both complex-valued. The real parts are shown in Figs.~2c and 4b and the imaginary parts are closely related to the ionization probability of the quantum orbit. For the simulations of the two-colour experiments, we considered the short quantum orbits from two optical cycles and summed their ionization probability amplitudes coherently. Their interference gives rise to the harmonic structures in the spectral domain. 

\section*{Data Availability}
\noindent The datasets generated and/or analysed during the current study will be published on the ETH data repository upon acceptance of the manuscript.

\section*{Code Availability}
\noindent The Octopus package used for the TDDFT calculations is publicly available. The remaining computer codes are available from the corresponding authors on reasonable request.

\begin{acknowledgments}
The authors thank Nirit Dudovich for insightful discussions, as well as Andreas Schneider and Mario Seiler for their contributions to the construction and improvements of the experiment. We acknowledge financial support from ETH Z\"{u}rich. ON gratefully acknowledges support from the Young Faculty Award from the National Quantum Science and Technology program of Israel's Council of Higher Education Planning and Budgeting Committee.
G.T. gratefully acknowledges support from the Alexander von Humboldt Foundation.
M.R. acknowledges funding by the European Union (ERC, QUADYMM, 101169761).
Z.Y. acknowledges funding by JSPS KAKENHI Grant-in-Aid for Scientific Research (A) (No. 25H00864).
\end{acknowledgments}

\section*{Author Contributions}
All authors participated in the discussion of the results and contributed to the manuscript.

\section*{Competing interests}
The authors declare no competing interests.
\bibliography{ref,attobib2}

\begin{figure*} [t!]
\includegraphics[width=\textwidth]{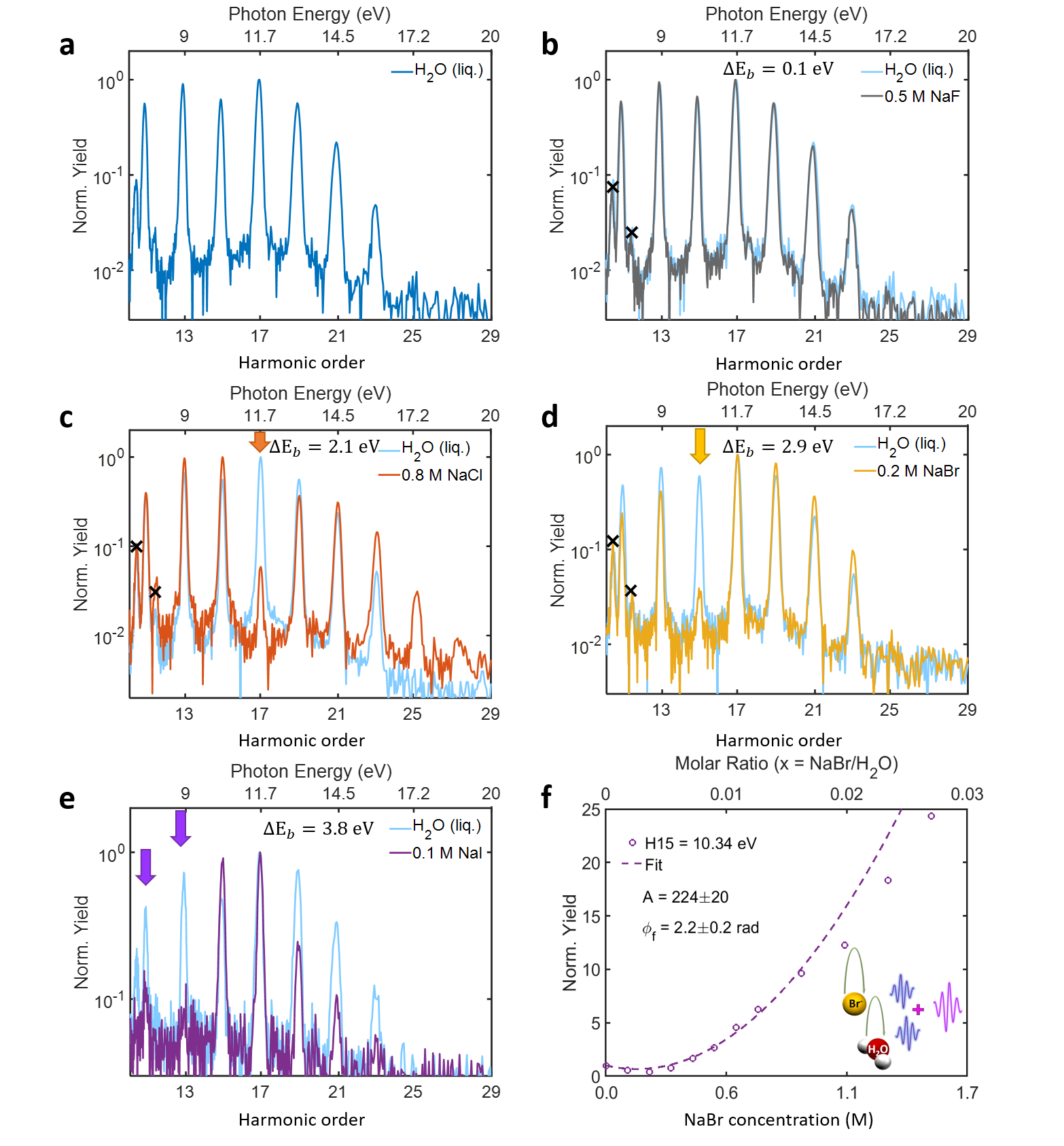}
{{\textbf{Extended Data Fig. 1}} 
 \textbf{Effect of the anions on HHG from aqueous solutions.} 
 \textbf{a} Normalized high-harmonic spectrum of pure H$_2$O, recorded with an 1800-nm driver at a peak intensity of 2.9$\times$10$^{13}$ W/cm$^2$. \textbf{b} 0.5 M NaF (grey) spectrum at a peak intensity of 2.9$\times$10$^{13}$ W/cm$^2$, overlayed on H$_2$O spectrum from (a). The spectra are nearly identical because the $E_{\rm b}$ of F$^-$ is very close to that of liquid H$_2$O, so no additional interference occurs. \textbf{c-e}, Normalized spectra of aqueous solutions of NaCl (0.8 M, orange), NaBr (0.2 M, yellow), and NaI (0.1 M, purple) solutions at 3.3$\times$10$^{13}$ W/cm$^2$, 3.6$\times$10$^{13}$ W/cm$^2$ and 2.9$\times$10$^{13}$ W/cm$^2$, respectively. As the anion changes from Cl$^-$ to I$^-$, $E_{\rm b}$ decreases relative to H$_2$O, increasing the $\Delta$E$_b$ and shifting the destructive interference condition to lower photon energies. This is observed as a clear suppression of harmonic yield at 11.7 eV in the case of Cl$^-_{\rm (aq)}$ (c, orange arrow), 10.3 eV in the case of Br$^-_{\rm (aq)}$ (d, yellow arrow) and 7.8 eV in the case of I$^-_{\rm (aq)}$ (e, purple arrows), consistent with the $E_{\rm b}$'s of 9.55 eV for Cl$^-$, 8.8 eV for Br$^-$, and 7.8 eV for I$^-$. Black crosses mark features that arise from the second diffraction order of the spectrometer grating. \textbf{f} Normalized yield of the most suppressed harmonic at $\sim$10.34 eV in the case of Br$^-_{\rm (aq)}$ as a function of salt concentration, recorded at $\sim$ 3$\times$10$^{13}$ W/cm$^2$.}
\end{figure*}

\begin{figure*} [t!]
\includegraphics[width=\textwidth]{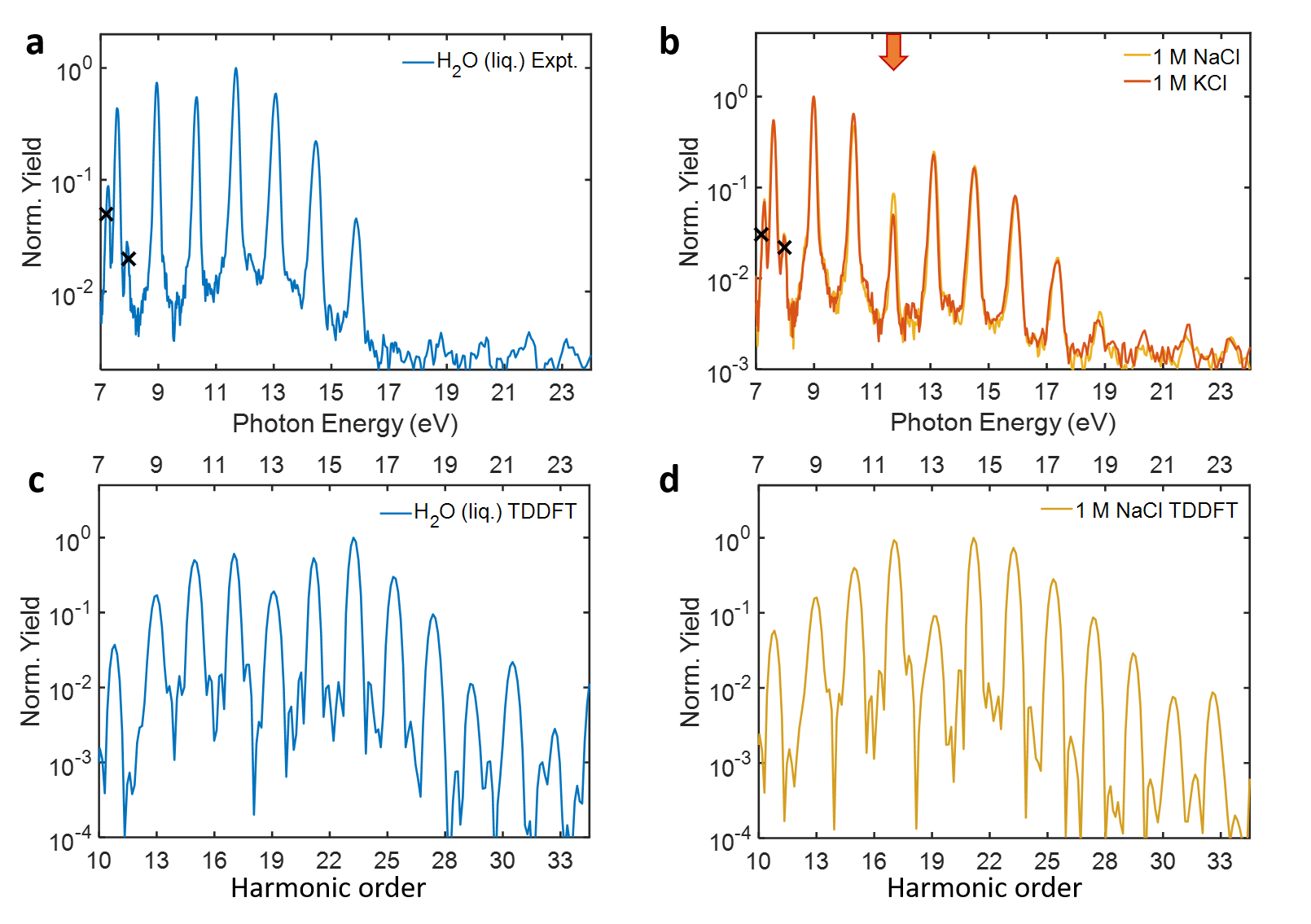}
{{\textbf{Extended Data Fig. 2}} 
\textbf{Effect of cations on HHG from aqueous solutions.}  
\textbf{a} Normalized high-harmonic spectra for pure liquid H$_2$O  at peak intensity of 3$\times$10$^{13}$ W/cm$^2$. 
\textbf{b} Normalized spectra of 1 M NaCl (yellow), and 1 M KCl (orange) solutions at peak intensity of 3.5$\times$10$^{13}$ W/cm$^2$. Here, the spectral profiles for different cations overlap, confirming that the cation species does not noticeably affect the harmonic emission. The differences between panel (a) and panel (b), especially the suppression of H17 (11.7 eV), arise from emission from the Cl$^-$ anion, which has a lower $E_{\rm b}$ than H$_2$O, leading to destructive interference effects that modulate the spectrum. Black crosses mark features that arise from the second diffraction order of the harmonics from the grating. (c) Calculated HHG spectrum of pure water from TDDFT at 2$\times$10$^{13}$ W/cm$^2$. (d) Calculated HHG spectrum of 1 M NaCl from TDDFT at 2$\times$10$^{13}$ W/cm$^2$. The driving wavelength in all panels was 1800 nm.}
\end{figure*}

\begin{figure*} [t!]
\centering
\includegraphics[width=\textwidth]{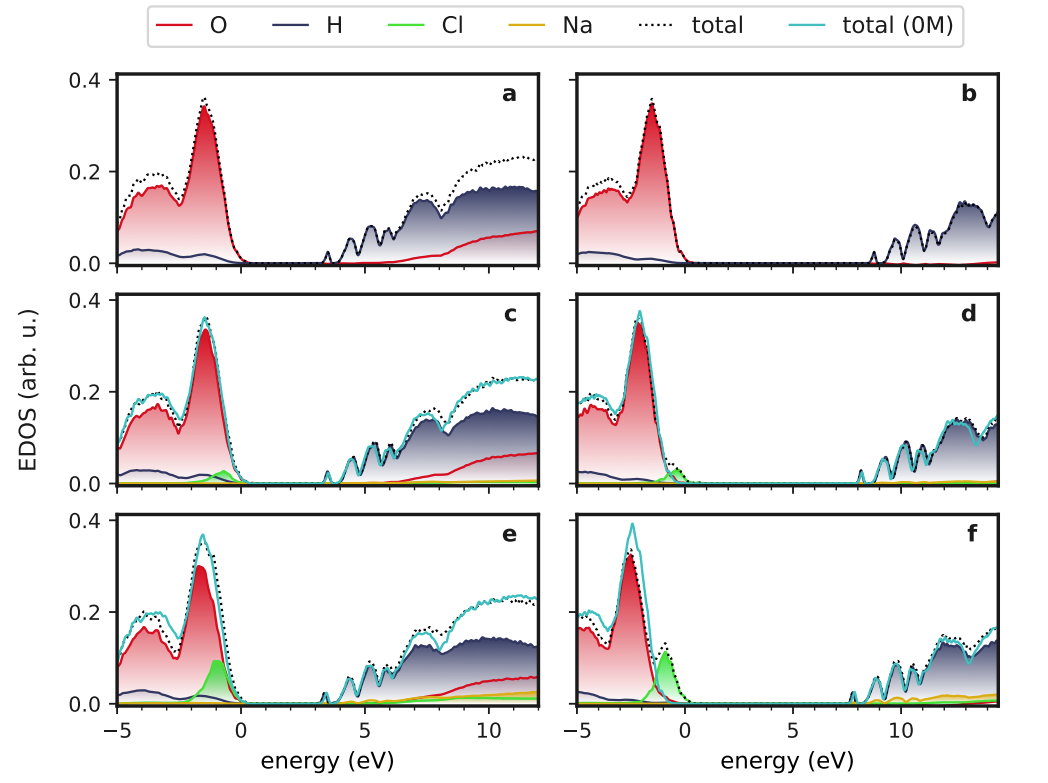}

{{\textbf{Extended Data Fig.~3}} 
\textbf{Electronic density of states of aqueous solutions.} 
Average electronic densities of states (EDOS) for pure water \textbf{(a,b)}, 1~M aqueous NaCl \textbf{(c,d)}, and 4~M aqueous NaCl \textbf{(e,f)}, obtained using the PBE functional (left) and PBE0 with 50\% exact exchange (right). For every subplot, the average was computed from the EDOSs of 50~trajectory snapshots, aligned by the mean oxygen core level. The energies were shifted so that the mean valence band maximum is at $\approx 0\,\text{eV}$. The EDOSs are normalized, such that the integral of the total density (dotted) from $-6\,\text{eV}$ to the band gap evaluates to 1. For comparison, the total EDOS of pure water [total (0M)] is shown in cyan alongside the EDOSs for the NaCl solutions.}
\end{figure*}

\begin{figure*} [t!]
\includegraphics[width=\textwidth]{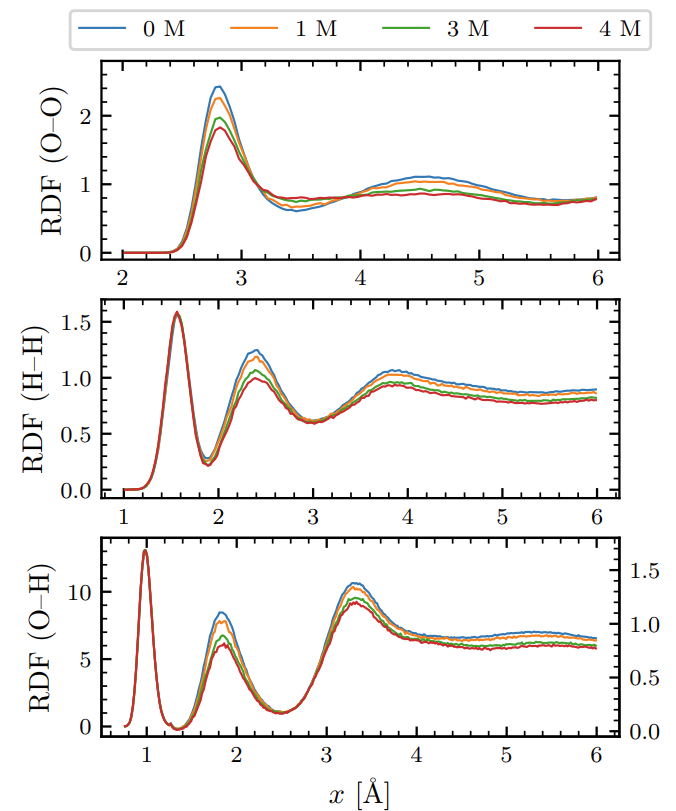}
{{\textbf{Extended Data Fig. 4}} 
\textbf{Radial distribution functions (RDFs) for aqueous NaCl solutions at varying concentrations.} 
RDFs for O–O (top), H–H (middle), and O–H (bottom) pairs derived from PIMD simulations are shown for salt concentrations of 0 M (pure water), 1 M, 3 M, and 4 M NaCl.}
\end{figure*}

\begin{figure*} [t!]
\includegraphics[width=\textwidth]{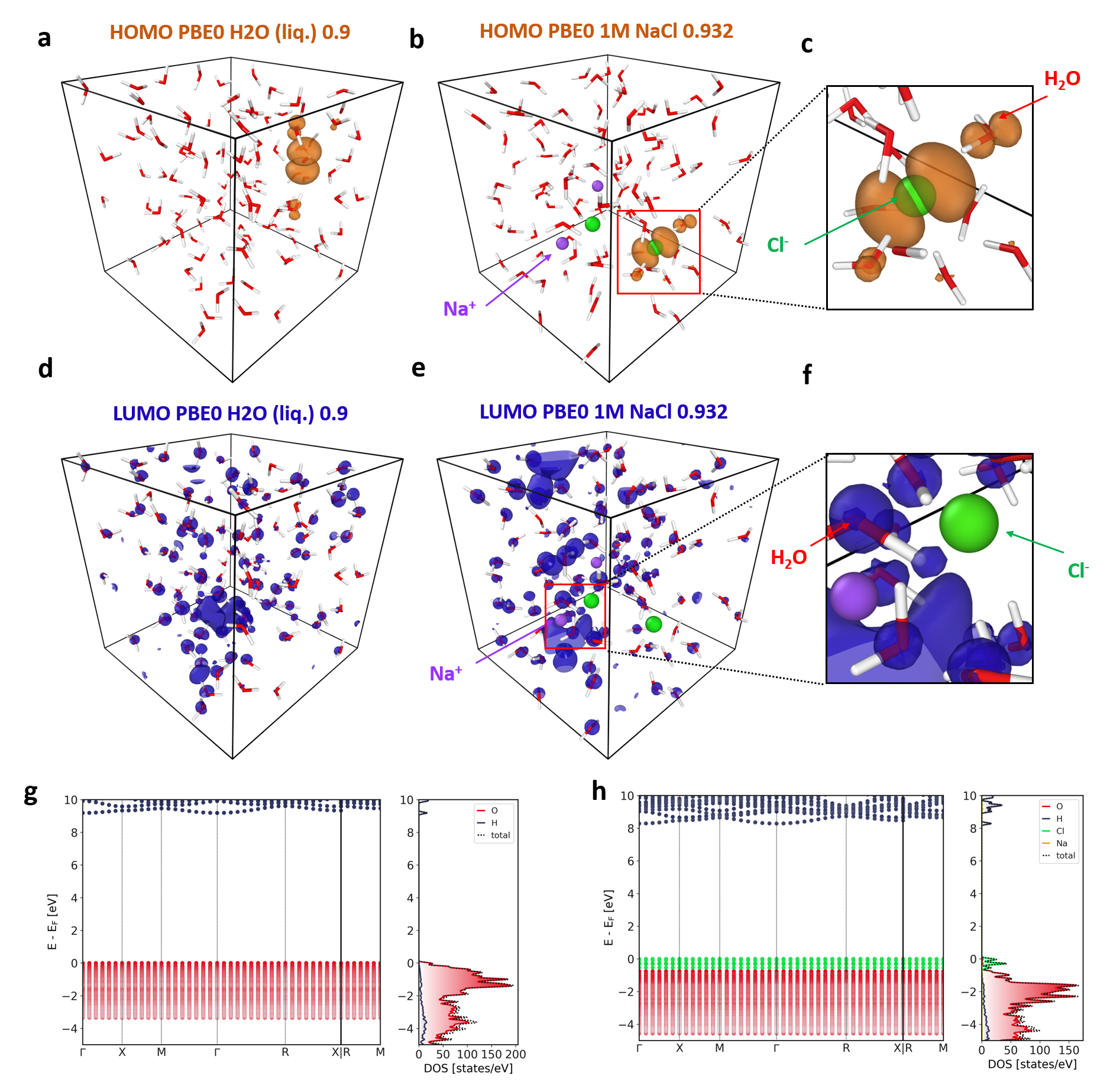}
{{\textbf{Extended Data Fig. 5}} 
\textbf{Orbitals and band structure of liquid water and aqueous NaCl solutions obtained with the PBE0 functional.}
\textbf{(a-f)} HOMO and LUMO of H$_2$O and a 1-M aqueous NaCl solution. Red dots correspond to oxygen atoms, white dots to hydrogen atoms, green dots to chloride anions and purple dots to sodium cations. \textbf{(g-h)} Band structures of H$_2$O (g) and a 1-M aqueous NaCl solution (h), for corresponding snapshots.}
\end{figure*}

\begin{figure*} [t!]
\includegraphics[width=\textwidth]{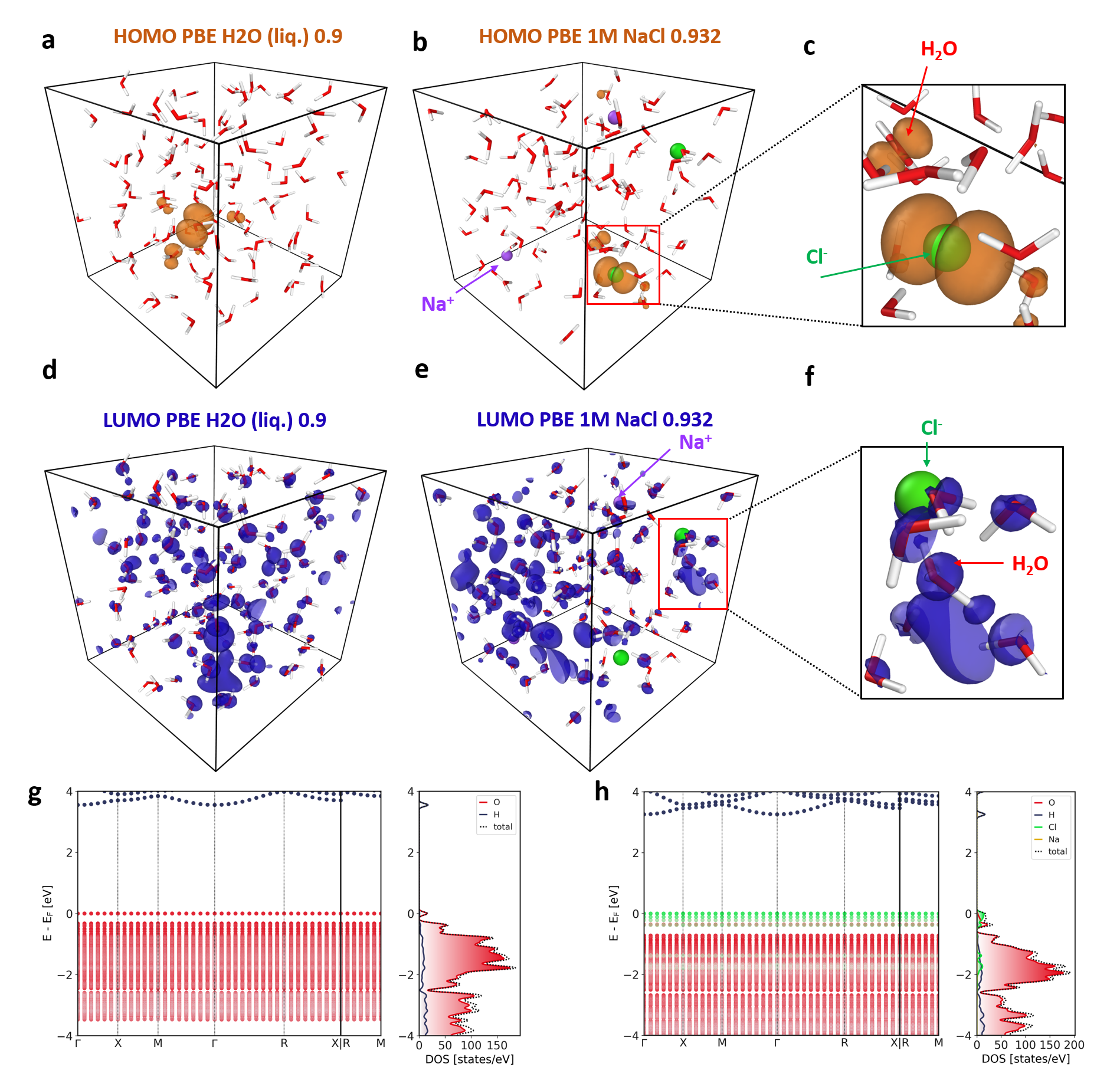}
{{\textbf{Extended Data Fig. 6}} 
\textbf{Orbitals and band structure of liquid water and aqueous NaCl solutions obtained with the PBE functional.}
\textbf{(a-f)} 
HOMO and LUMO of H$_2$O and a 1-M aqueous NaCl solution. Red dots correspond to oxygen atoms, white dots to hydrogen atoms, green dots to chloride anions and purple dots to sodium cations. \textbf{(g-h)}  Band structures of H$_2$O (g) and a 1-M aqueous NaCl solution (h), for corresponding snapshots.}
\end{figure*}

\begin{figure*} [t!]
\includegraphics[width=\textwidth]{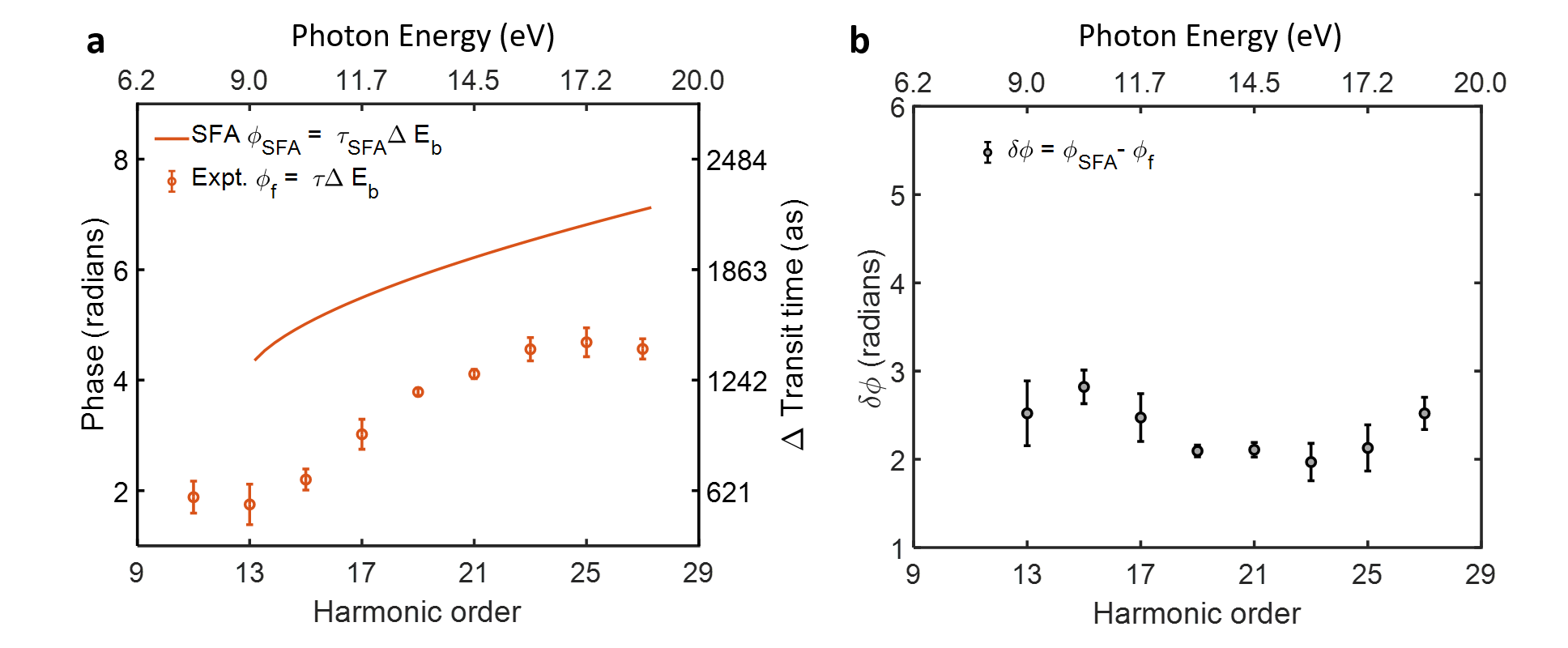}
{{\textbf{Extended Data Fig. 7}}
\textbf{Role of the dipole phase in passive interferometry}
\textbf{(a)} Comparison of the relative phase $\phi_{\rm f}$ determined by passive interferometry (symbols) and the corresponding variation in transit time with the transit time from a quantum-orbit calculation (solid line) described in the Methods Section, using a driving wavelength of 1800 nm, a peak intensity of 3.3 $\times$ 10$^{13}$ W/cm$^2$ and $E_b=\,$8.83~eV (bandgap of liquid water). The nearly energy-independent offset is interpreted as the signature of the dipole phase.
\textbf{(b)} Phase difference ($\delta\phi=\phi_{\rm SFA}-\phi_{\rm f}$), which is interpreted as the difference of the dipole phases of solute (Cl$^-$) and solvent (H$_2$O).}
\end{figure*}

\end{document}